# Search for RPV Scalar Leptons at Tevatron

**Shan-Huei Chuang***†
*University of Wisconsin - Madison, USA*
*E-mail:* `schuang@fnal.gov`

We reviewed CDF and DØ searches for $R$-parity violation supersymmetry in leptons involved final states using up to $344 \pm 21$ pb$^{-1}$ Tevatron Run II data of $p\bar{p}$ collisions at $\sqrt{s} = 1.96$ TeV. All the results were in good agreement with the Standard Model expectations. No evidence of new physics was observed. However, owing to the improvement on detectors, energy and luminosity from Run I to Run II, the limits for the existence of $R$-parity violation supersymmetry have been greatly advanced.

*Speaker.
†On Behalf of the CDF and DØ Collaborations

## 1. Introduction

Although the Standard Model (SM) has been remarkably successful in describing presently known phenomena there are arguments to consider it as a low-energy effective, hence the construction of its extensions to high energies. Supersymmetry (SUSY) is the most appealing SM extension. Mainly motivated by the cancellation of quadratic divergences in scalar boson mass loop corrections, SUSY postulates such a symmetry that every fermion has a bosonic superpartner and every boson has a fermionic superpartner [1]. The $R$ parity is defined as $R \equiv (-1)^{3B+L+2J}$. $R = \pm 1$ for $\substack{\text{the Standard Model}\\ \text{most SUSY models}}$. SUSY by itself does not impose $R$ parity conservation. In a minimal $R$ parity violation ($\not{R}$) SUSY model the gauge invariance of the minimal supersymmetric extension of Standard Model (MSSM) allows the following additional Yukawa couplings in the Lagrangian:

$$\mathscr{L}_{\not{R}} \equiv [\lambda_{ijk} L_i L_j \bar{E}_k + \lambda'_{ijk} L_i Q_j \bar{D}_k + \lambda''_{ijk} \bar{D}_i \bar{D}_j \bar{U}_k] \tag{1.1}$$

where $i, j, k$ are family indices [2]. The $\lambda$ and $\lambda'$ terms violate lepton number (LNV) and the $\lambda''$ terms violate baryon number (BNV). Proton decay via squark $\tilde{d}_2$ forbids the co-existence of LNV and BNV but not either one. In analogy to the standard Yukawa couplings one expects a hierarchical structure among the $\not{R}$ couplings and hence the lightest superparticle (LSP) to decay dominantly into one of the 45 $\not{R}$ channels.

## 2. Search for $R$-parity Violation Decay of Sneutrino

CDF searched in $344 \pm 21$ pb$^{-1}$ Run II data for high mass resonance decaying into an electron plus a muon and interpreted the results in the context of $\not{R}\,\lambda'_{311}$ production and $\lambda_{132}$ decay of $d\bar{d} \to \tilde{\nu}_\tau \to e\mu$ [3]. The high $p_t$ $e\mu$ signature let in little SM backgrounds. The search used high $p_t$ lepton triggers and selected events with $N_e \geq 1$ and $N_\mu \geq 1$, requiring each lepton to have $p_t > 20$ GeV and be isolated and each lepton pair to be close at beam axis with $|\Delta z_0| < 5$ cm and have opposite signs of charge. Signal and SM backgrounds (except QCD) were modeled using PYTHIA (jet data) with the $h^0_{SM}$ particle being a functional substitute for $\tilde{\nu}_\tau$. See Figure 1. Uncertainties on acceptance included 9.2% due to MC generator, 3.2% due to momentum resolution and 2.4% due to parton distribution function; on backgrounds the largest was the 16% from QCD. Observing no excess in data 95% CL limits of $\tilde{\nu}_\tau$ production cross-section as well as of $\lambda'_{311}$ and of $\lambda_{132}$ were set. See Figure 2.

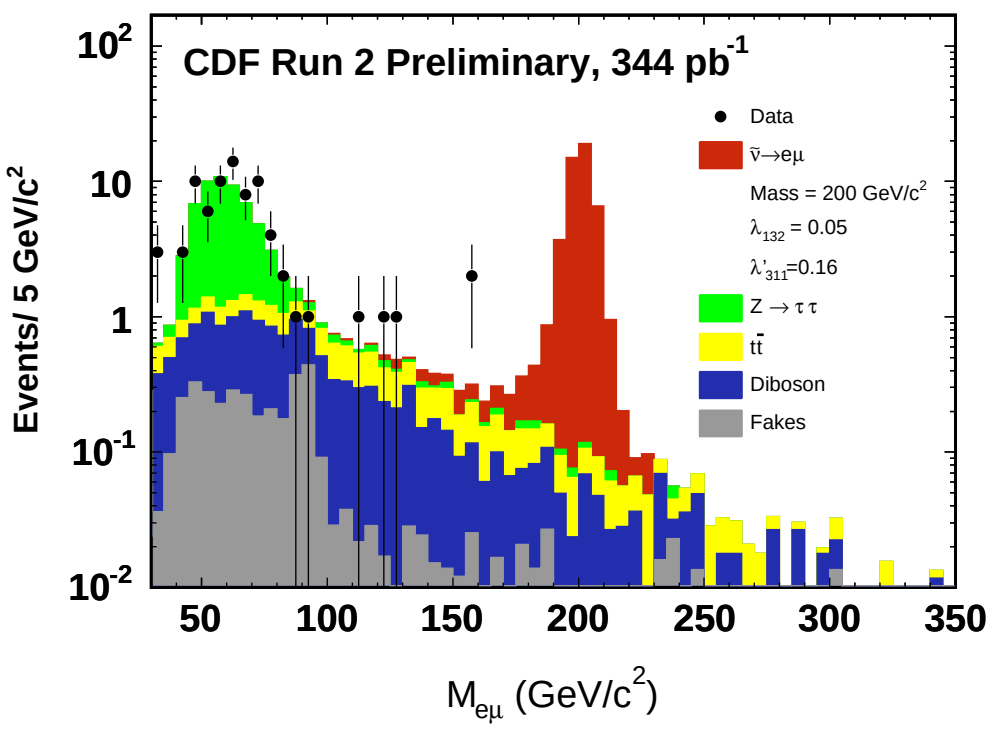

**Figure 1:** The mass spectrum of data observation vs SM expectation, showing good agreement. The $\tilde{\nu}_\tau$ signal contribution is hypothetical.

## 3. Search for $\not{R}$ Production and Decay of Slepton via the $LQ\bar{D}$ Coupling $\lambda'_{211}$

DØ searched in $154 \pm 10$ pb$^{-1}$ Run II data for the resonant production of $\tilde{\mu}$ which leads to a final state with two muons and two jets and the advantage of all the masses reconstructable,

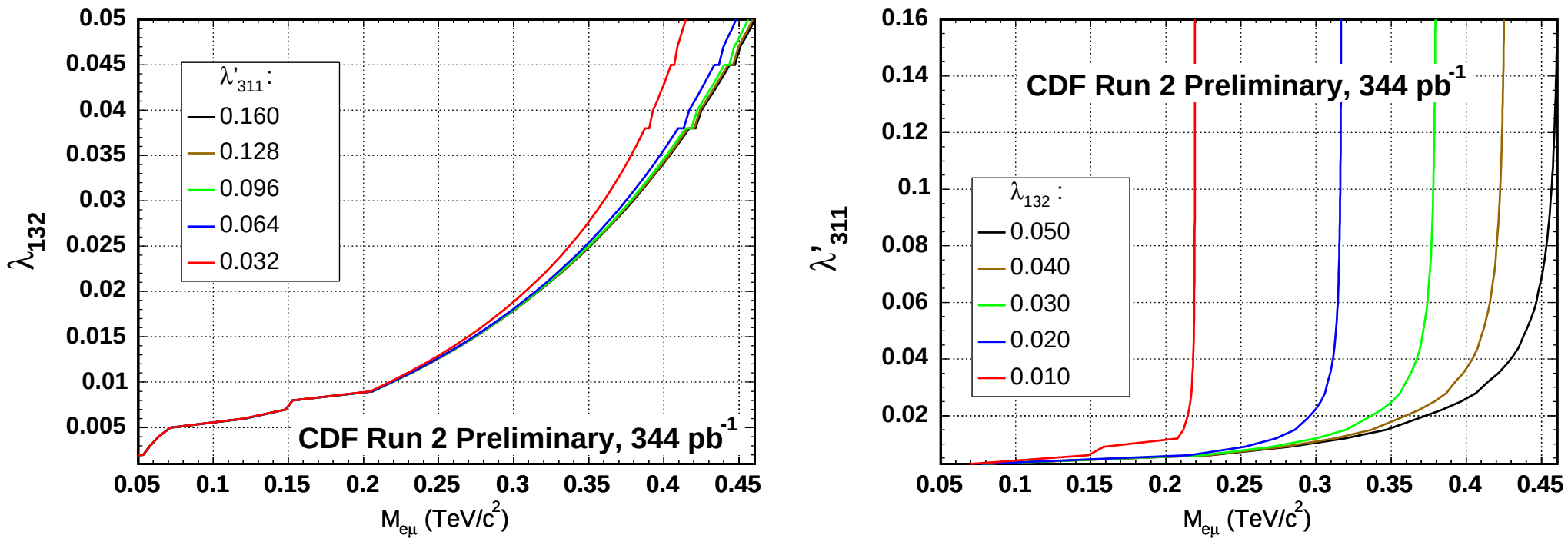

**Figure 2:** The upper limits of $\lambda_{132}$ for fixed $\lambda'_{311}$ (left) and of $\lambda'_{311}$ for fixed $\lambda_{132}$ (right) as a function of $m_{\tilde{\nu}_\tau}$.

Figure 3 (left), assuming $\lambda'_{211}$ is the dominant $\not{R}$ coupling [4]. The search used dimuon triggers and selected events with S/$\sqrt{\text{B}}$-optimized mass dependent cuts: $\not{E}_t < 60$ GeV, 2 isolated muons with $p_t^{\{\mu_1,\,\mu_2\}} > \{21.25+0.1875(m_{\tilde{\mu}}-m_{\tilde{\chi}_1^0}),10\}$ GeV and $p_t^{\mu_1}+p_t^{\mu_2} > 60$ GeV, 2 jets in $|\eta| < 2$ and $E_t^{\{j_1,\,j_2\}} > \{25,15\}$ GeV, $m_{\tilde{\chi}_1^0}^{\text{MC}}-40 < m_{\tilde{\chi}_1^0} < m_{\tilde{\chi}_1^0}^{\text{MC}}+20$ GeV, $|m_{\tilde{\mu}}^{\text{MC}}-m_{\tilde{\mu}}| < 0.2m_{\tilde{\mu}}^{\text{MC}}$, rejecting $Z$ with $|m_{\mu\mu}-91| < m_{\tilde{\chi}_1^0}/9+5$ GeV and suppressing QCD by requiring large angle between any two final-state objects. Signal was modeled with SUSYGEN, scanning over 25 $\{m_{\tilde{\mu}},m_{\tilde{\chi}_1^0}\}$ points with $\lambda'_{211}=0.07, A_0=0, \tan\beta=2$ and the higgsino mass mixing parameter $\mu<0$. The only non-negligible background $Z$+2$j$ was modeled with PYTHIA. By far the largest uncertainty $\sim 30\%$ was due to jet energy scale. Reasonable agreement was found for all points, e.g. expected : observed $= 1.1\pm0.4 : 2$ events. Observing no excess in data 95% CL limits of $\tilde{\mu}$ production cross-section and of $\lambda'_{211}$ were set, as shown in Figure 3 (middle and right).

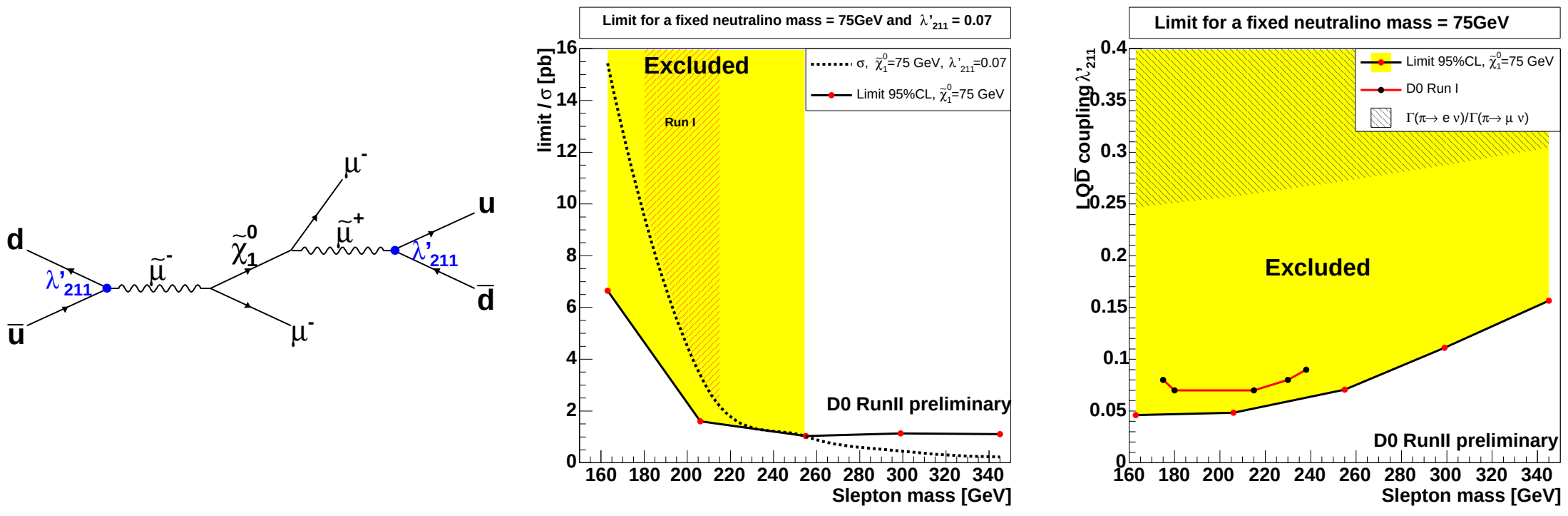

**Figure 3:** Feynman diagram for $\not{R}\,\tilde{\mu}$ production and leptonic $\tilde{\chi}_1^0$ decay (hadronic not drawn) via $\lambda'_{211}$. The $\mu$ or quark directly from the $\tilde{\chi}_1^0$ decay is expected to have low $p_t$ since its sparticle sister is highly virtual (left). The $\tilde{\mu}$ production cross-section limits (middle) and the $\lambda'_{211}$ limits (right) as a function of $m_{\tilde{\mu}}$.

## 4. Search for RPV Decay of Chargino/Neutralino into $\geq 3$ Lepton Final State

DØ has searched for $\not{R}$ leptonic decays of $\tilde{\chi}_1^0$ via $\lambda_{121}$, $\lambda_{122}$, $\lambda_{133}$ into final states that contain $ee\ell$, $\mu\mu\ell$, $ee\tau_h$ where $\ell \in \{e,\mu\}$ separately [5, 6, 7]. The searches commonly considered the sce-nario in which, given the $\not{R}$ couplings have been set significantly smaller than the gauge couplings by low-energy experiments [8], most SUSY processes conserve $R$ parity and only LSP can $\not{R}$-decay into SM particles. Sparticles can only be pair produced at Tevatron [1]. In the processes of $p\bar{p} \rightarrow$ SP-pair ... $\rightarrow$ LSP-pair ... $\stackrel{\not{R}}{\rightarrow} \ell\ell\nu\ell\ell\nu$ ... at least 4 leptons are in final state. Event selections went commonly as follow: isolated high or medium $p_t$ dielectron or dimuon with invariant mass outside $\Upsilon$ and $Z$ regions, an additional low $p_t$ lepton (see Figure 4), and significant $\not{E}_t$ presence to reduce QCD background; the 4th lepton was not required to increase signal acceptances. Signals were modeled with SUSYGEN, scanning over $140 < m_{1/2} < 280$ GeV. SM backgrounds (except QCD), of which the largest was Drell-Yan, were estimated with PYTHIA (jet data). Uncertainties on acceptances were driven by the 6% on luminosities; on backgrounds 60% of total were from DY and QCD. Results and related SUSY parameters are summarized in Table 1.

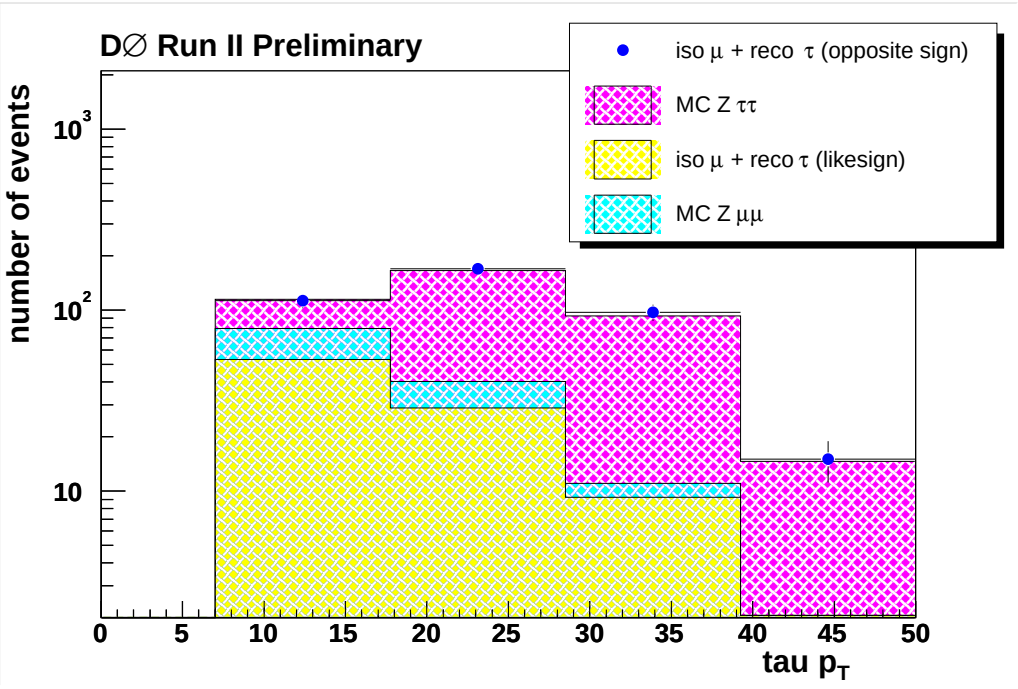

**Figure 4:** The $p_t$ spectrum of hadronically decayed tauon ($\tau_h$) candidates after neural network identification.

| | $\mathscr{L}_{\text{int}}$ (pb$^{-1}$) | $\lambda$ constraint | $\tan\beta$ | $m_0$ | $N_{\text{expected}} : N_{\text{observed}}$ | $m^{\mu<0}_{\tilde{\chi}^0_1}$ | $m^{\mu<0}_{\tilde{\chi}^\pm_1}$ | $m^{\mu>0}_{\tilde{\chi}^0_1}$ | $m^{\mu>0}_{\tilde{\chi}^\pm_1}$ |
|---|---|---|---|---|---|---|---|---|---|
| $ee\ell$ | $238 \pm 16$ | $\lambda_{121}$=0.010 | 5 | 250 | $0.45 \pm 0.43 : 0$ | 95 | 181 | 97 | 183 |
| $\mu\mu\ell$ | $160 \pm 10$ | $\lambda_{122}$=0.001 | 5 | 250 | $0.63 \pm 1.93 : 2$ | 84 | 160 | 90 | 165 |
| $ee\tau_h$ | $199 \pm 13$ | $\lambda_{133}$=0.003 | 10 | 80 | $1.04 \pm 1.42 : 0$ | | | 66 | 118 |

**Table 1:** Summary of the $\not{R}$ SUSY trilepton searches, all in good agreement with the SM expectations. $A_0 = 0$. The 95% CL lower $m_{\tilde{\chi}^0_1}$ or $m_{\tilde{\chi}^\pm_1}$ limits (GeV) for $\mu > 0$ or $\mu < 0$ are shown in the last four columns.

## 5. Conclusion

As CDF and DØ have searched so far there is no evidence of $\not{R}$ SUSY yet. The limits of $R$-parity violation supersymmetry have been either newly set or pushed way further.